\documentclass[12pt,preprint]{aastex}

\shorttitle{Radio Galaxies at z = 1.1 to 3.8}
\shortauthors{Steinbring et al.}

\input epsf
\def\plotone#1{\centering \leavevmode
\epsfxsize=1.0\columnwidth \epsfbox{#1}}

\def\plotonefiveeigths#1{\centering \leavevmode
\epsfxsize=0.625\columnwidth \epsfbox{#1}}
\def\plotonetwothirds#1{\centering \leavevmode
\epsfxsize=0.666\columnwidth \epsfbox{#1}}
\def\plotonethreequarters#1{\centering \leavevmode
\epsfxsize=0.75\columnwidth \epsfbox{#1}}

\begin{document}

\title{Radio Galaxies at z = 1.1 to 3.8: Adaptive-Optics Imaging and 
Archival Hubble Space Telescope Data}

\author{E. Steinbring\altaffilmark{1}}
\affil{Center for Adaptive Optics, University of California, Santa Cruz, CA
95064}
\email{steinb@ucolick.org}

\and

\author{D. Crampton\altaffilmark{1} and J. B. Hutchings\altaffilmark{1}}
\affil{Dominion Astrophysical Observatory, Herzberg Institute of Astrophysics,
National Research Council of Canada, 5071 West Saanich Road, Victoria, BC 
V8X 4M6, Canada}

\altaffiltext{1} {Visiting Astronomers, Canada-France-Hawaii Telescope,
operated by the National Research Council of Canada, the Centre de la
Recherche Scientifique de France, and the University of Hawaii}

\begin{abstract}
We have undertaken a program of high-resolution imaging of high-redshift radio
galaxies (HzRGs) using adaptive optics on the
Canada-France-Hawaii Telescope. We report on deep imaging in $J$, $H$,
and $K$ bands of 6 HzRGs in the redshift range 1.1 to 3.8. At these
redshifts, near-infrared bandpasses sample the rest-frame visible galaxian
light. The radio galaxy is resolved in all the fields and is generally
elongated along the axis of the radio lobes. These images are compared to
archival Hubble Space Telescope Wide-Field Planetary Camera 2 optical observations of
the  same fields and show the HzRG morphology in rest-frame ultraviolet and
visible light is generally very similar: a string of bright compact knots. Furthermore, this sample - although very small - suggests the colors of the knots are consistent with light from young stellar populations. If true, a plausible explanation is that these objects are being assembled by mergers at high redshift. 
\end{abstract}


\section{Introduction}\label{introduction}

The range of redshifts of the known radio galaxies (RGs) spans from $z=0.06$ for \objectname{Cygnus A} to well above $z=3$. For example, \objectname{4C +41.17} at $z=3.79$, \objectname{6C 0140+326} at $z=4.41$, and 
\objectname{TN J0924-2201} at $z=5.19$. It is well established that the most
powerful low redshift radio sources are associated with giant elliptical
galaxies, often the brightest cluster galaxies.
The optical morphology of powerful RGs with redshifts greater than $\sim 0.6$
is dramatically different from that of those of lower redshift. These
high-redshift radio galaxies (HzRGs) are clumpy and very irregular and tend 
to be elongated along the axis of their radio lobes, a characteristic known
as the `alignment effect' \citep{Chambers1987, McCarthy1987}. 
The best optical imaging of HzRGs has been obtained
with the Hubble Space Telescope (HST) Wide-Field Planetary Camera 2 (WFPC2).
These data include $\sim 100$ objects mostly from the
3C and 4C catalogs and have revealed very complex rest-frame UV
morphologies down to the resolution limit of $\approx 0.1${\arcsec}
\citep{Longair1995, Best1996, Best1997, Chambers1996, McCarthy1997, Pentericci1999}.
These images reveal the immediate neighbourhood of the AGN host to be strings of
knots with separations on the order of a few kpc aligned along the radio axis.
These are embedded in a diffuse emission region with projected scales of
$\sim 50$ kpc. Typically, several faint field objects
within $\sim 100$ kpc projected radius from the host are also found.

A still unanswered question about the stellar population of HzRGs is
whether it is young or old. Since the highest redshifts for RGs are in the 3 - 5
range, one might assume that they are very young systems since, for a universe
with $q_0=0.5$, the look-back time is about 90\% of the age of the universe.
A popular theory presented by \citet{Lilly1984} \citep{Lilly1988, Lilly1989} 
is that most of the star formation in these systems took place in an initial burst at $z_{\rm formation}\sim 5-10$ and that minor star-formation episodes afterwords produce the dramatic morphologies in rest-frame UV light. This
`old galaxy $+$ burst' model is reasonably successful in
matching the $K-z$ relation for RGs but it is not the only way to account for
the rest-frame UV light. The high incidence of double and multiple component galaxies with separations of a few kpc suggests that these may be in the process of merging in the direction defined by the radio axis \citep{Djorgovski1987, West1991, West1994}. Star formation  could be induced by the shocking of the intergalactic medium by jets along the radio axis \citep{Chambers1987, McCarthy1987, DeYoung1989, Rees1989, Begelman1989, Chambers1990, Daly1990}. The high linear polarization of some galaxies suggests that scattering of light from the active galactic nucleus (AGN) by electrons or dust may be the dominant source of emission \citep{diSeregoAlighieri1989, Tadhunter1992, diSeregoAlighieri1993, Januzzi1995, Cimatti1996, Cimatti1997}. Also, nebular thermal continuum associated with ionized gas is an important contributor for some objects \citep{Dickson1995, Stockton1996}. 

The rest-frame UV morphologies of HzRGs may be dramatic but in order to study
galaxy evolution in HzRGs one should follow the mature stellar populations. If
these populations are mapping out the structure of the `true' galaxy one might
well ask how this structure changes in time. Do these galaxies appear as
elliptical galaxies from early epochs to the present or, if not, how do they
evolve? Deep, high-resolution NIR imaging allows the study of HzRGs to be extended to
the most distant ones known. It permits the
investigation of the rest-frame-visible properties of RGs over a large range in
redshift and enables the discrimination between opposing viewpoints on HzRG
formation and evolution - mature, red, passively evolving ellipticals or young,
bursting irregulars - by permitting studies of the morphology of the stellar
populations of HzRG environments. 
Work on imaging HzRGs in the NIR using the HST
have been hampered by the failure of the Near-Infrared Camera and Multi-Object
Spectrometer (NICMOS) in January 1999. From a sample of 19 $1.7<z<3.2$ RGs with 
NICMOS $H$-band images \citet{Pentericci2001} suggest that the lower redshift galaxies are well fit by de Vaucouleurs profiles. \citet{Zirm1999} have obtained
images of a further 11 RGs and also suggest the NIR morphologies are
consistent with dynamically relaxed elliptical host galaxies.

The work discussed here involves a program of deep imaging of HzRGs with the
Canada-France-Hawaii Telescope (CFHT) Probing the Universe with Enhanced Optics (PUEO) AO system that was designed to provide similar depth and resolution in the NIR
as that provided by HST in the optical \citep{Steinbring1998}. The aim of the
study was to discover if
the compact knots found in rest-frame UV had corresponding rest-frame-visible
counterparts or if the stellar HzRG morphology was more like that of an evolved
elliptical galaxy. We present CFHT PUEO observations of 6 HzRGs with 
$1.1 < z < 3.8$. Deep HST WFPC2 images of the same fields had already 
been obtained by others and were publically available from the Archive, but 
in two cases these data were unpublished. In Section~\ref{sample_selection} we discuss sample selection and in
Section~\ref{data} the combined HST and CFHT dataset and its reduction is outlined. Descriptions
of the individual fields and comparison of the WFPC2 and PUEO images will follow
in Section~\ref{descriptions}. Analysis of the morphology and colors of the
RGs in the sample is presented in Sections~\ref{analysis}; conclusions
follow in Section~\ref{conclusions}.

\section{Sample Selection}\label{sample_selection}

Ideally, a study of the stellar environments of HzRGs should have a sample that
covers a wide range in redshift and radio brightness. The major discovery
surveys of RGs contain several hundred sources each. Thus, there are thousands
of RGs from which to choose a sample that should satisfy these criteria. 

The difficulty is, however, that very few of these galaxies will satisfy
the selection criteria for observation with a natural guide-star AO
instrument such as the CFHT PUEO. The major problem is that a suitable
guide-star must not only be bright but in close proximity on the sky to the RG.
Thus, in effect, the selection criterion for the RG sample is purely
based on the proximity of a guide star. 
This may yield a very small sample but at least it is assured to be chosen
without bias for a particular physical trait. Radio galaxy surveys select
objects based on radio flux which is not at all dependent on the projected
proximity of stars. One might suggest that their identification with an optical
counterpart does prohibit ones with bright stars very close to, or, worse,
superimposed on, the galaxy. But this will not, of course, correlate with radio
brightness or redshift or any other intrinsic parameter of the RG sample either.

Possible RG targets were chosen in the following fashion. First,
all objects catalogued by the NASA Extragalactic Database as an RG with
$z>1.0$ and visible from the CFHT during our observing runs were found. That is,
these were galaxies that were identified with an optical component. This
yielded a target list including hundreds of candidates. Next, the United States
Naval Observatory Guide-Star Catalogue was used to determine all of those targets within 45{\arcsec} of an $R<15$ star. This reduced the target list
to only 25 members.
Finally, of the targets remaining, only those with archival HST WFPC2 images
were included. The last step was applied to ensure that high-resolution optical
observations would be available for all the targets. This reduced the sample to
12 members; all but one of which were from either of the 3C and 4C Catalogues of radio sources.

\section{Data and Reductions}\label{data}

We obtained CFHT PUEO data during observing runs in June 1997 and January 1998.
A journal of the observations is given in Table~\ref{table_cfht_pueo}. See
\citet{Rigaut1998} for a discussion of the PUEO instrument and its performance.
In all observations a $12<R<15$ guide-star was used to guide PUEO. During
the 1997 run only the $256\times 256$ pixel Montreal NIR Camera 
(MONICA) was available. Its
poor sensitivity hampered our efforts to obtain deep exposures of the RG fields.
The small field of view (FOV) (9{\arcsec} $\times$ 9{\arcsec} with
0.0347{\arcsec} pixels) meant that neither the guide-star nor any star suitable
for determining the point-spread function (PSF) was included in the target
field. The $1024\times 1024$ pixel NIR imager KIR was used in our 1998 run 
and its greater
sensitivity ($\sim 4\times$ that of MONICA) provided much deeper exposures.
The larger format (0.0350{\arcsec} pixels giving a field of 36{\arcsec} $\times$
36{\arcsec}) also included either the guide-star or some other stars suitable
for PSF analysis. We also used these stars to register the PUEO and WFPC2
images.

For the observations reported here the natural seeing was between 0.5{\arcsec}
and 0.9{\arcsec} with good transparency. We made observations primarily through
standard $H$ and $K$ filters. A few observations through standard $J$ were also
made. A $K'$ filter would have been preferred because of the lower resultant
sky-noise but none was available for our runs. We were unable to obtain any $K$
data during the 1997 run due to problems with the filter-wheel.
Exposures of 300 s each were made in a non-repeating rectangular dither pattern
with 1{\arcsec} steps for MONICA and 4{\arcsec} steps for KIR.
An $H$-band image of a single object, \objectname{3C 356}, was obtained in 
June 1997 with MONICA. Images of 5 more fields were acquired in 
January 1998 with KIR.
 
\subsection{Archival Hubble Space Telescope Data}\label{archival_data}

The archival WFPC2 data were obtained through the Canadian Astronomy Data
Centre which is operated by the National Research Council of Canada
Herzberg Institute of Astrophysics. The images were combined using
standard Space Telescope Science-Database Analysis System tasks and
rotated to the proper orientation based on information contained in the image
headers. 
A list of the observations is given in Table~\ref{table_hst_wfpc2}.
In all cases the MONICA or KIR data were resampled to match the pixel
sampling of the WFPC2 data. No smoothing was done prior to resampling.

\subsection{Calibration}\label{calibration}

Photometric standard stars from the United Kingdom Infrared Telescope
catalog were observed. Star FS 30 was used as a calibrator in June 1997 and the
star FS 21 was used in January 1998. For the archival HST data the photometric
zero-points and first-order transformations to the standard Johnson filter set
were found in the WFPC2 Instrument Handbook \citep{Biretta1996}.

In the remainder of the analysis the HST WFPC2 F675W and F702W filters are
assumed to be approximately equivalent to $R$ and the F785LP and F814W filters
also roughly equivalent to $I$. Only a first order transformation from the local
to standard Johnson filter set has been made for these and the CFHT data.

For both the WFPC2 and KIR images the PSF FHWM at the galaxy position was
determined by measuring a bright but unsaturated star in the field. For WFPC2
images this was the star nearest the RG and for KIR it was a star at a similar
offset from the guide-star. In the case of \objectname{3C 294} no such PSF star was available and no guide-star calibration observations were made but the galaxy is within 9{\arcsec} of a $R=12$ magnitude star. We therefore 
followed these observations with imaging of a crowded stellar field in the globular cluster \objectname{NGC 4147}, also
using a $R=12$ guide-star. An
isolated star roughly 9{\arcsec} from the guide-star in NGC 4147 was used as 
the PSF for the scientific observation.
 
For MONICA no PSF stars were available due to its small FOV. However, for the
MONICA observations of \objectname{3C 356} exposures of the guide-star were available. These
were taken immediately before each set of galaxy exposures and were combined to
determine an average on-axis PSF. The PSF reconstruction technique of \citet{Veran1996} was also employed to determine an on-axis PSF and this 
agreed with the observed one. The PSF at the {\it target} position was then determined using calibration observations of a crowded stellar field and the technique discussed in \citet{Steinbring2001}. Observations of the globular
cluster \objectname{M 5} were made. A mosaic pattern was used to obtain a
roughly 12{\arcsec} wide by 30{\arcsec} long strip with the guide-star at one
end. The mosaic was composed of 6 individual pointings of the telescope. At each
position a dithering pattern was used to build up an image. Since the
guide-star appeared in only one of the component images and time lapse of
several minutes occured between each pointing a danger with this method was that
the PSF would change during the construction of the mosaic. This did not appear
to be a problem because stars in overlapping regions in the mosaic were almost
identical. In each overlapping region the FWHM of a particular star in both
frames was not different by more than 5\%.
The \objectname{M 5} calibration field was then deconvolved with its
guide-star. The Lucy-Richardson algorithm was employed and this process was
permitted to continue until all the flux in the guide-star image was
concentrated into roughly one pixel. This produced an image of deconvolved
stars. These stars are also the `kernels' necessary to degrade an image of the
on-axis PSF to one at an off-axis position. Next, the deconvolved image was convolved with the on-axis PSF for the
scientific observation of \objectname{3C 356}. This produced an off-axis PSF correct for both
the seeing conditions and anisoplanatic effects during the \objectname{3C 356} observations.
   
The FWHM of the PSFs were determined by fitting them with a Gaussian using the
IRAF task IMEXAM. The results are presented in 
Table~\ref{table_image_resolution}. 
All the galaxies were resolved, typically at resolutions of 0.20{\arcsec} to
0.25{\arcsec} FWHM. The per-pixel $S/N$ for bright knots within the RGs is
approximately 10 for both the CFHT and HST data. For the fainter knots and
field objects this drops to $S/N\approx3$.

\subsection{Photometry}\label{photometry}

The CFHT images were searched by eye
to find stars, knots, and faint objects in the field. 
Also, the software package SExtractor \citep{Bertin1996} was used to find objects in the field but it was determined that this generally selected
the same objects as those found by eye. Each RG field was corrected for galactic extinction with values from the
extinction map of \citet{Schlegel1998}. Coordinate lists of objects in the
fields were generated and synthetic aperture photometry was performed for all
of these objects in the CFHT and HST images.

The IRAF task APPHOT was used employing a 3.0{\arcsec} aperture on all objects
in the field. This aperture was large enough to encircle all the flux for each
of the targets. For the RGs themselves rectangular apertures were determined
that separated the individual knots without overlap. This was always done in the
band corresponding most closely to the rest-frame visible for each galaxy and the same apertures were maintained
for the other bands. We defined a knot as a bright region of
connected pixels that was separated from any other region by a trough of pixels
with flux lower than half of the maximum brightness. This task was accomplished
by analyzing the image of the RG with a display interface that permitted the
flux in any pixel to be displayed when the cursor was pointed at it. By scanning
the cursor over pixels radiating from a local maximum in the image it was
straightforward to determine which peaks were connected and which were isolated.
The results are shown in Table~\ref{table_photometry}.
The subcomponents of the RG are labeled with a lower-case letter and we assume $H_0=70$ km ${\rm s}^{-1}$ ${\rm Mpc}$ and $q_0=0.5$ when quoting their $H$ magnitudes. Other field objects are indictated by a number.
The quoted 1-${\sigma}$ photometric errors are due to Poisson statistics from sky-flux. In some cases the formal errors were somewhat less than 0.1 magnitude
but uncertainties in flat-fielding, extinction, and photometric zero-points 
suggest a lower limit on the photometric uncertainty of 0.1 magnitude.

Another consideration is emission-line contamination of the broadband fluxes. These emission lines are the products of ionization by the hidden AGN and it is
important to know if they will seriously interfere with what should be
measurements of stellar continuum light. The following are strong lines
redshifted into the observed bandpasses: \ion{Si}{4}~$\lambda$1403, 
\ion{C}{4}~$\lambda$1549, \ion{C}{3}]~$\lambda$1909, 
[\ion{Ne}{5}]~$\lambda$3426, [\ion{O}{2}]~$\lambda$3727, 
[\ion{Ne}{3}]~$\lambda$3869, H${\beta}$, [\ion{O}{3}]~$\lambda$5007, 
[\ion{N}{2}]~$\lambda$6548, 6583, and H${\alpha}$. If a spectrum was available the contamination was estimated by comparing the
FWHM of the filter bandpass and the equivalent width of the emission lines. 
No NIR spectra were available for two of the targets so the equivalent
widths here were estimated from a composite RG spectrum constructed from
observations of galaxies with $0.1<z<3$ \citep{McCarthy1993}. 
The results are shown in Table~\ref{table_emission_line}. The estimated
contamination is always less than 20\% and suggests that emission lines will not have a significant effect on the photometric results.
It should be noted that even with the advantage of an observed spectrum this method only gives an estimate for the galaxy as a whole since emission-line strengths will not be uniform among its components. However, in our
photometry the light of each RG was divided among a just few sub-apertures
and there is no evidence for strong discord over these spatial scales since
the colors generally agree within photometric uncertainty.

\section{Descriptions of Each Field}\label{descriptions}

The CFHT and HST images follow. First, 
either our $H$ or $K$-band AO image is shown. Each image has north up and east left with right ascension and declination given in J2000 coordinates. Stars are indicated by an `S', the
guide-star by `GS', and the calibration star by `PSF'. Some well-resolved galaxies much brighter than the RG were detected and these are indicated by `FG' for foreground galaxy. The peaks of emission of the radio lobes are denoted by `X's. Enlarged fields showing the RG itself are 
given for each of the available bands including the HST data. Each field is 4{\arcsec} $\times$ 4{\arcsec} with north up and east left.

\subsection{3C 356}\label{3c356}

This is a $z=1.08$ radio bright (11.3 Jy at 178 MHz) RG.  
This RG has a very extended double radio lobe morphology. The projected separation of the lobes is just over an arcminute or about 250 kpc in physical size. A high-resolution
(0.18{\arcsec} FWHM) VLA radio map at 8.4 GHz was obtained by Best et al. (1997)
and indicates two unresolved radio sources located roughly midway along the radio axis. 

\begin{figure}
\plotonefiveeigths{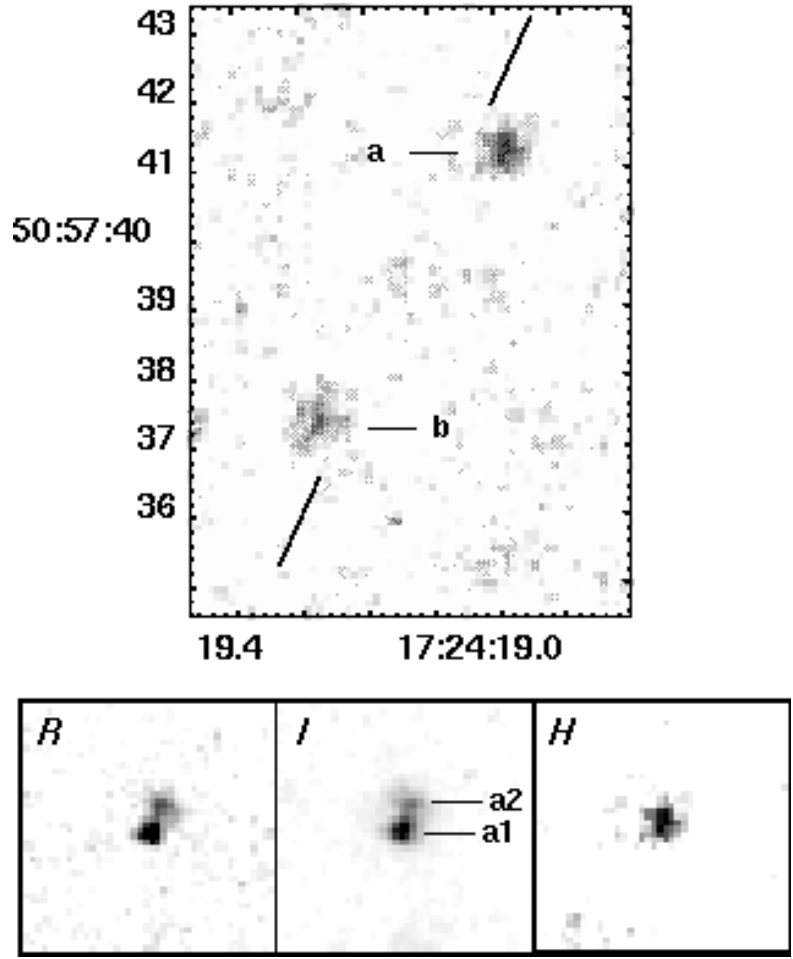}
\caption{Images of the 3C 356 field. North is up and east left. Right ascension
and declination are given in J2000 coordinates. Our $H$-band image is shown
at the top. Below are images of the RG in the 3C~356
field. The FOV for each of the bottom three panels is
4{\arcsec}$\times$4{\arcsec}. See text for details.} 
\label{figure_3c356}
\end{figure}

Previous non-AO ground-based NIR observations had been obtained
by \citet{Eisenhardt1990}, \citet{Rigler1992}, and Best et al. (1997). These images indicated that the central radio sources correspond to the two brightest rest-frame-optical components. Both components were, however, unresolved.
These two objects, `a' and `b', are resolved in our $H$-band imaging and have about the same brightness. 
The top panel of Figure~\ref{figure_3c356} shows component `a' to be elongated along roughly a north-south axis while object `b' is more diffuse.
 
The WFPC2 $R$ and $I$ images of component `a' are shown along with
our $H$ image in the bottom three panels of Figure~\ref{figure_3c356}. The HST data are discussed in detail by \citet{McCarthy1997} and Best et al. (1997), and show component `a' to be `dumbbell' shaped. McCarthy et al. and Best et al. suggest that since the optical morphology of `a' is more typically that of RGs of the 3C Catalogue it is the AGN host. This identification is confirmed by our NIR image. 
The marked difference in morphology between `a' and `b' is also exhibited
in our $H$-band image. Furthermore, the axis of elongation of component `a'  
matches that seen in the HST images.

Keck Telescope optical spectropolarimetry by \citet{Cimatti1997} reveals that
both components are polarized in rest-frame UV (observed 4000-9000 \AA) with
$P\approx 8$\%. This continuum polarization rises towards the blue, reaching
$P\approx15$\% for the northern component. Observations with the UK Infra-Red
Telescope by \citet{Leyshon1998} show that both components are also probably polarized in
rest-frame visible light (observed $K$-band) at the 10\% level as well. These
observations suggest an obscured AGN as a likely source of scattered non-stellar light. This further suggests that there might be pollution of our photometry with AGN light. Some contamination had also been suggested by the analysis
of emission lines in Section~\ref{photometry}.
     
\subsection{3C 230}\label{3c230}

This is a $z=1.487$ radio-bright (19.2 Jy at 178 MHz) RG. \citet{Rhee1996}
obtained VLA radio observations of \objectname{3C 230} at 4.8 GHz and 8.4 GHz.
These moderate resolution (0.4{\arcsec} FWHM) maps indicate that it has a double lobed morphology but no core object is detected.

\begin{figure}
\plotone{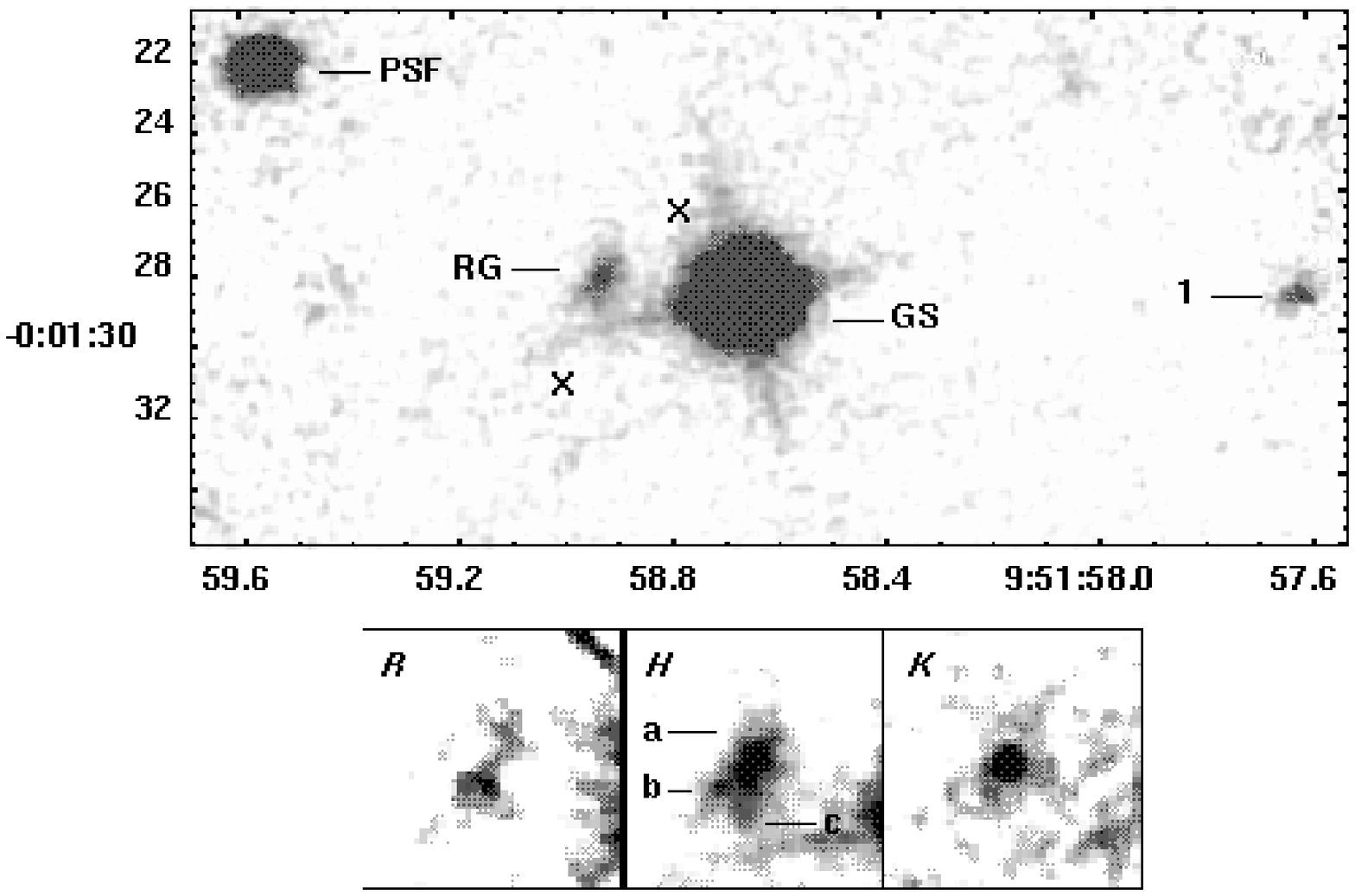}
\caption{Same as Figure~\ref{figure_3c356} for the 3C 230 field. See text for
details.}
\label{figure_3c230}
\end{figure}

Previous attempts to obtain ground-based NIR imaging of this field were hampered
due to the proximity of the bright ($R=15$) star 4{\arcsec} to the west \citep{Hammer1990}. This
star provided an excellent guide for AO observations. Our $H$-band image is 
shown in the top panel of Figure~\ref{figure_3c230}. The RG itself is extended and composed of 3 knots. They are aligned roughly parallel with the axis of the radio lobes; offset by about 1{\arcsec} to the northeast. No knots are detected at the positions of the radio lobes. The bright object at the western edge of the field is probably a foreground galaxy. 

The WFPC2 $R$ data were obtained by others as part of a snapshot survey of 3C radio galaxies \citep{Sparks1995}. These data are now public
but were previously unpublished. The $R$-band image of the RG is shown
along with our NIR data in the bottom three panels of Figure~\ref{figure_3c230}. The $R$-band image also shows the RG to be elongated along the axis of the radio lobes. Interestingly, the fainter central region of the RG in $R$ (southeast
of knot `a') is also the location of the brightest component in $H$. This is especially apparent in
the $K$ image and is perhaps due to obscuration by dust in the galaxy.

\subsection{3C 68.2}\label{3c68.2}

This is a radio bright (10.0 Jy at 178 MHz) RG at $z=1.575$. A high-resolution
(0.18{\arcsec} FWHM) VLA radio map at 8.4 GHz was obtained by 
Best et al. (1997).
The radio emission has the classic extended double lobed
morphology with an axis running roughly southeast to northwest. No core is
detected to a 3-$\sigma$ flux limit of 0.13 mJy.

\begin{figure}
\plotonethreequarters{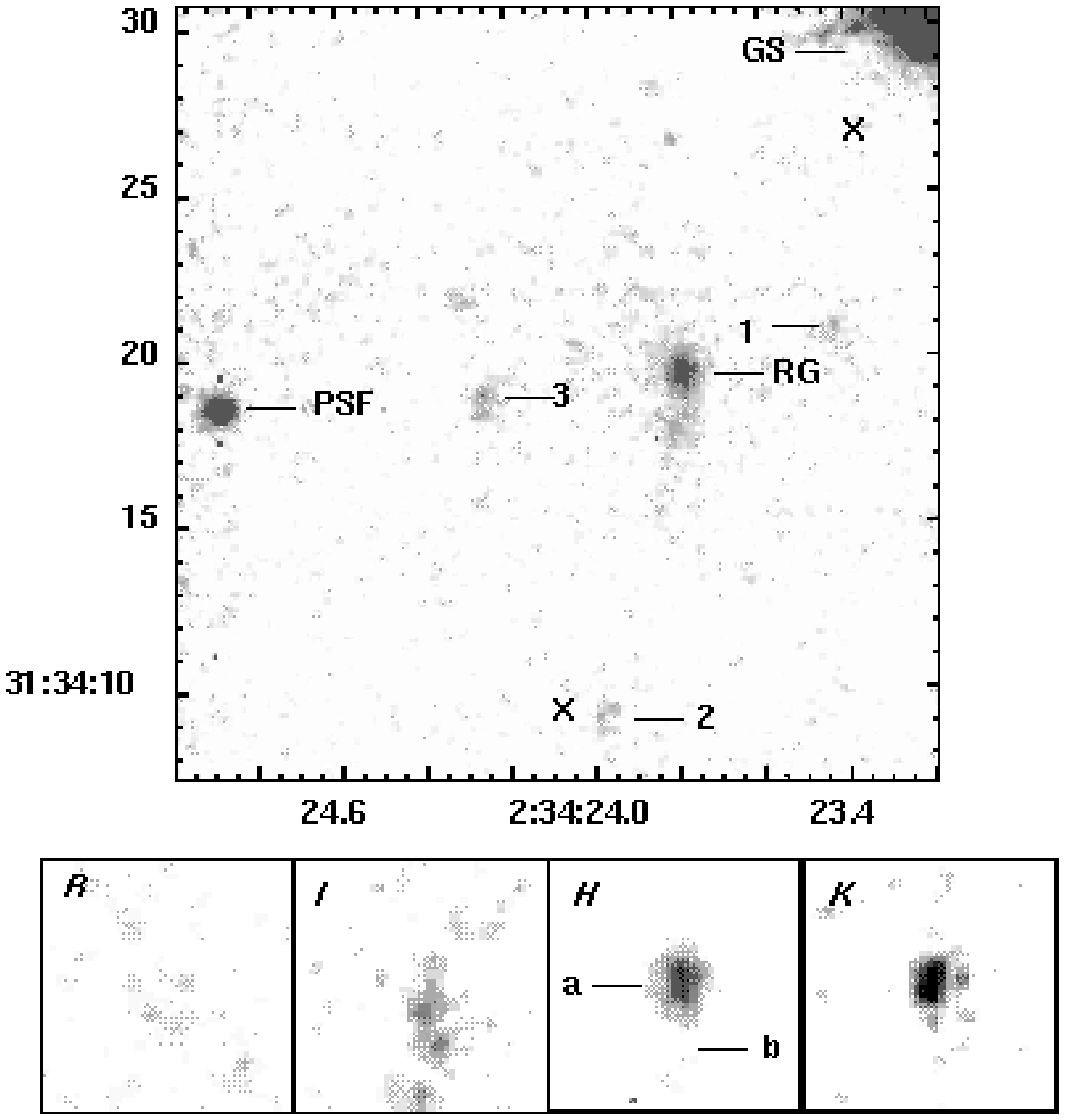}
\caption{Same as Figure~\ref{figure_3c356} for the 3C 68.2 field. See text for details.}
\label{figure_3c68.2}
\end{figure}

Previous NIR observations had been obtained by Best et al. (1997). The $K$-band image was acquired with $\sim1${\arcsec}
seeing and showed only that the RG was extended and roughly aligned with the radio axis. Our $H$-band image appears at the top of Figure~\ref{figure_3c68.2} and shows that the galaxy is extended and misaligned with the radio axis by 20 degrees. The RG
itself is composed of 2 components separated by 1.5{\arcsec} - denoted by `a' and `b' - in the $H$ image. Object `b' is more diffuse than `a' and although it is faint, there
is some indication that it trails 2 or 3 arcseconds towards the southern radio lobe. Other objects are detected in the field. Field object `1' is diffuse
and lies roughly along the axis of the radio emission. Field object `2' is within 2{\arcsec} of the southern radio hotspot. It was not detected in
the Best et al. $K$ image but in our $H$ image it appears to be more compact than `1'. We also detect a third field object, labelled `3', that lies roughly 6{\arcsec} east of the RG. The images of both `2' and `3' are extended and, furthermore, they share a common axis of elongation parallel to that of the RG.

The WFPC2 data are discussed in detail by Best et al. (1997). The $I$-band
image of the RG itself is displayed along with our $H$ and $K$ data in the bottom three panels of Figure~\ref{figure_3c68.2}. These show 
the RG to be more complex in the HST image. Three knots
correspond to the extended component `a' seen in the $H$-band image, and there
is also a knot corresponding to component `b'. The faint tail seen in
the $H$ image extends southward 2{\arcsec} from this knot. Best et al.
suggest that component `2' (labelled `f' by them) may be at the same
redshift as the RG and that its blue color is due to scattered light from
an obscured AGN or to star-formation induced by shocks associated with
the nearby radio lobe. Our NIR data reveals that components `2' and `3'
have similar colors, each is aligned with the RG, and thus both may
be related to it.

\subsection{3C 294}\label{3c294}

This is a  $z=1.78$ radio bright (10.0 Jy at 178 MHz) RG. A moderate resolution
(0.4{\arcsec} FWHM) VLA map at 5.0 GHz has been obtained by \citet{McCarthy1990} and indicates that \objectname{3C 294} has a double lobed
morphology along an axis from northeast to southwest. An unresolved core is
also detected. 

\begin{figure}
\plotonetwothirds{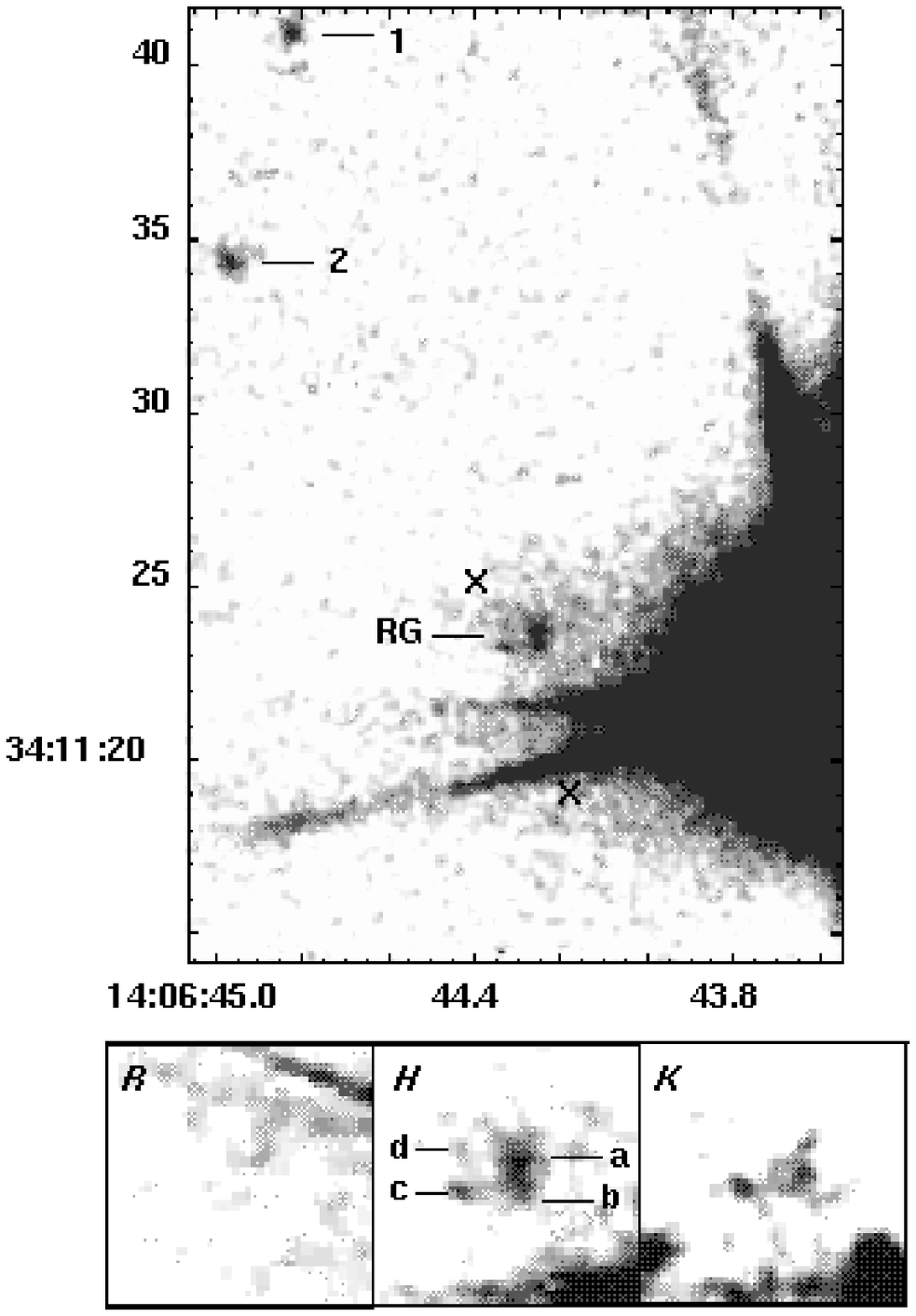}
\caption{Same as Figure~\ref{figure_3c356} for the 3C 294 field. See text
for details.}
\label{figure_3c294}
\end{figure}

Attempts to obtain non-AO ground-based NIR images of the RG have been hampered by the
bright ($R=12$) star 9{\arcsec} to the west \citep{McCarthy1990}. This proximity was an asset for AO imaging. Our $H$ image is shown in the top panel of Figure~\ref{figure_3c294}. The RG is in the centre of the field. The two other bright objects in the northeast seem to be
positioned along the axis of the radio emission although they are far outside the northeastern radio lobe. The colors of these objects are also similar to those of the RG.
The $H$ and $K$ images of the RG itself (bottom panels of Figure~\ref{figure_3c294}) are not aligned with the radio axis. 
The main component is elongated along a north-south axis
with a companion about 2{\arcsec} to the east of the southern tip. 
The brightest knot in $K$ lies 0.2{\arcsec} west of the main structure
in $H$, which may indicate some reddening for the RG.
The unresolved core (flux density 0.56 mJy) from the \citet{McCarthy1990} 
5.0 GHz radio map does not correspond directly with any of the knots in our images. It
is at a position 0.7{\arcsec} due south of the peak of knot `b'. The uncertainty
in this position is about 0.2{\arcsec}. This is because in this case the
guide-star is double with a separation of 0.13{\arcsec} and the positional error
quoted by McCarthy et al. for their radio map is 0.05{\arcsec}.

The bright potential guide-star near this galaxy has lead others to obtain AO
observations of \objectname{3C 294}. Similar resolution images were obtained by \citet{Stockton1999} using the
University of Hawaii AO system on the CFHT. Their $K'$ image is of comparable depth
and shows the RG to be very similar to our $K$ image. They also place the
\citet{McCarthy1990} radio core at or near the southern edge of the RG. Based on this location, they suggest that the most plausible scenario
is that of small dusty clumps being illuminated by an obscured AGN. 
A higher resolution
($\sim 0.05${\arcsec} FWHM) $H$-band image of similar depth to ours has been obtained with the Keck II AO system by \citet{Quirrenbach2001} which further resolves components `a' and `b' into several compact knots. Their derived
position for the radio core corresponds to component `c', which
remains unresolved in their image. This leads them to conclude that a
more plausible explanation for the NIR morphology is that of an
ongoing merger event, with the AGN located in the less massive of two galaxies.

The WFPC2 $R$ image had previously been viewed as a nondetection \citep{McCarthy1997}. 
It is included in the bottom three panels of Figure~\ref{figure_3c294}.
With the position of the galaxy now known in the NIR it is possible that the visible counterpart
in the WFPC2 image is discernible. We suggest that this is the faint nebulosity partially obscured by the diffraction spike of the star.
Several tests were carried out to ensure that it is not an artifact.
The WFPC2 data consist of 8 equal integrations taken in either of two independent camera orientations. Different sub-sets of the data were co-added using a variety of cosmic-ray and bad-pixel-rejection methods. 
The object is faint but persists independently of the different
processing. This indicates that the RG is certainly red, with $R-H\approx4.4$, 
and thus parts of it may be obscured by dust. 

\subsection{TXS 0828+193}\label{txs0828+193}

This is a radio bright (0.549 Jy at 365 MHz) RG at $z=2.57$. Moderate resolution
(1.5{\arcsec} FWHM) VLA radio maps at 1.5 GHz by \citet{Roettgering1994}
indicate that it has a double lobed morphology. A high-resolution
(0.25{\arcsec} FWHM) map at 8.2 GHz by \citet{Carilli1997} reveals a central core.   

\begin{figure}
\plotonetwothirds{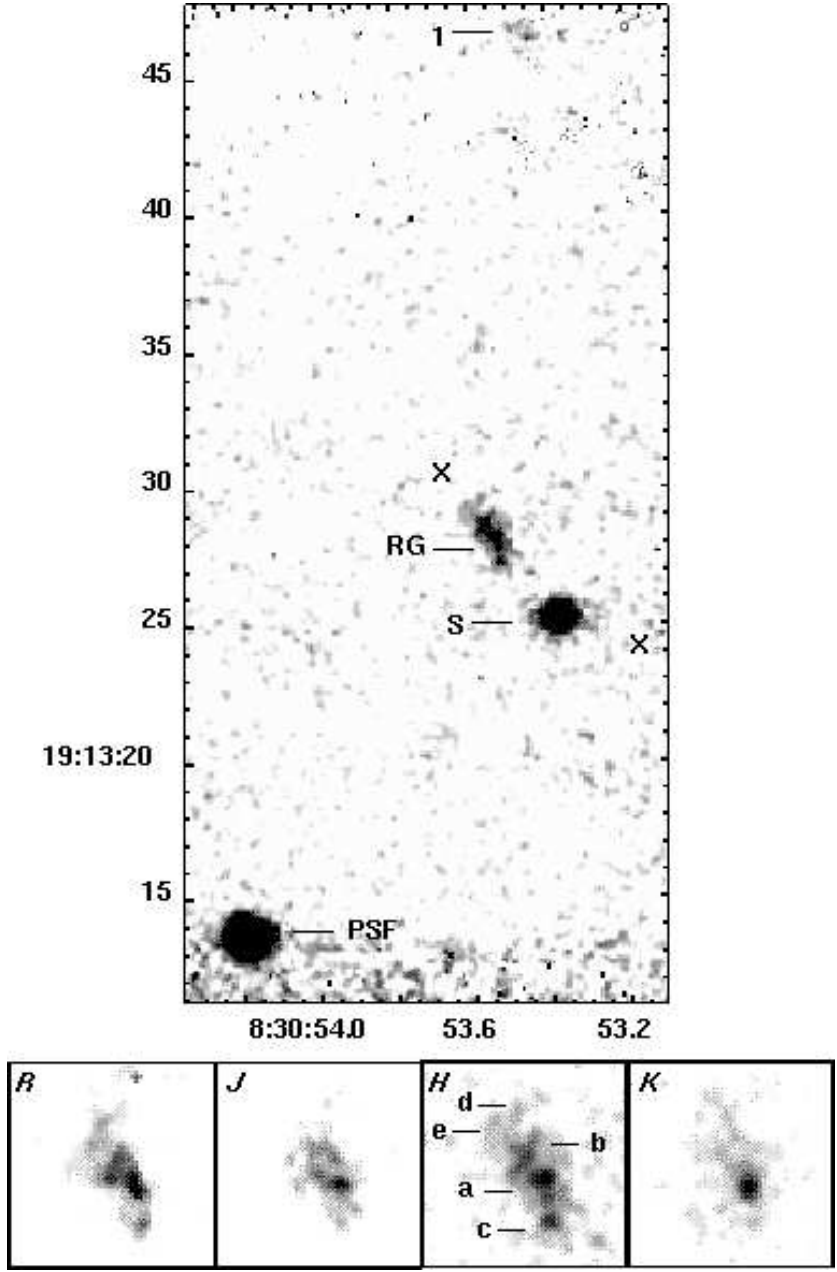}
\caption{Same as Figure~\ref{figure_3c356} for the TXS 0828+193 field. See
text for details.}
\label{figure_txs0828+193}
\end{figure}

Non-AO NIR images were obtained by \citet{Knopp1997} but were
obtained with 1{\arcsec} seeing and thus the RG is only marginally resolved.
The RG is spectacular in our $H$ image (top panel of Figure~\ref{figure_txs0828+193}). It is composed of approximately
5 knots which are aligned with the radio axis. The bottom panels
of Figure~\ref{figure_txs0828+193} show it to be less complicated
in the $K$ image, with a dominant knot that does not correspond exactly
with the brightest emission in the $J$ and $H$ images. Since $K$ corresponds roughly to rest-frame $R$ for this redshift the bluer light may be surpressed which might indicate internal reddening for this galaxy. 
There is a faint object
18{\arcsec} to the north which is misaligned with the radio axis but has colors very similar to the RG. The bright object 3{\arcsec} southwest of the RG is clearly pointlike in our images and NIR spectroscopy by \citet{Evans1998} reveals no emission-line
or absorption features. This may therefore be an unrelated object
along the line of sight. 

The HST $R$ image is discussed in detail by \citet{Pentericci1999} and
is displayed along with our $J$, $H$, and $K$ images in the bottom four
panels of Figure~\ref{figure_txs0828+193}. Pentericci et al. identify the
radio core with the brightest optical component. They suggest that the 
triangular morphology of the RG indicates an ionization cone
with a large fraction of the rest-frame UV light originating from scattered
light from an obscured AGN. Our NIR images show a striking correspondence
between the rest-frame UV and visible morphologies. The brightest knot in $R$
is also dominant at NIR wavelenghts. Components `b' and `c' in $R$ correspond
exactly in $J$ and $H$. The fainter structure 
at the northern end of the $R$ image corresponds to knots `e' and `d' in
our $H$ image. Whatever its origin, the emission in rest-frame UV is therefore likely to be from the source that dominates in visible light.

\subsection{4C +41.17}\label{4c+41.17}

At $z=3.80$ this is the highest-redshift object in the sample. Indeed it is one
of the RGs with highest known redshift. It is a radio loud object with a 178 MHz
flux of 2.7 Jy. Radio maps at several frequencies between 1.5 GHz and 15.0 GHz
have been obtained with the VLA by \citet{Chambers1990b}, \citet{Carilli1994}, and \citet{Chambers1996}. The axis of the radio emission is roughly northeast to southwest and is double lobed. An elongated central source and a compact core
are also detected in the higher resolution (0.2{\arcsec} FWHM) maps. 

\begin{figure}
\plotonethreequarters{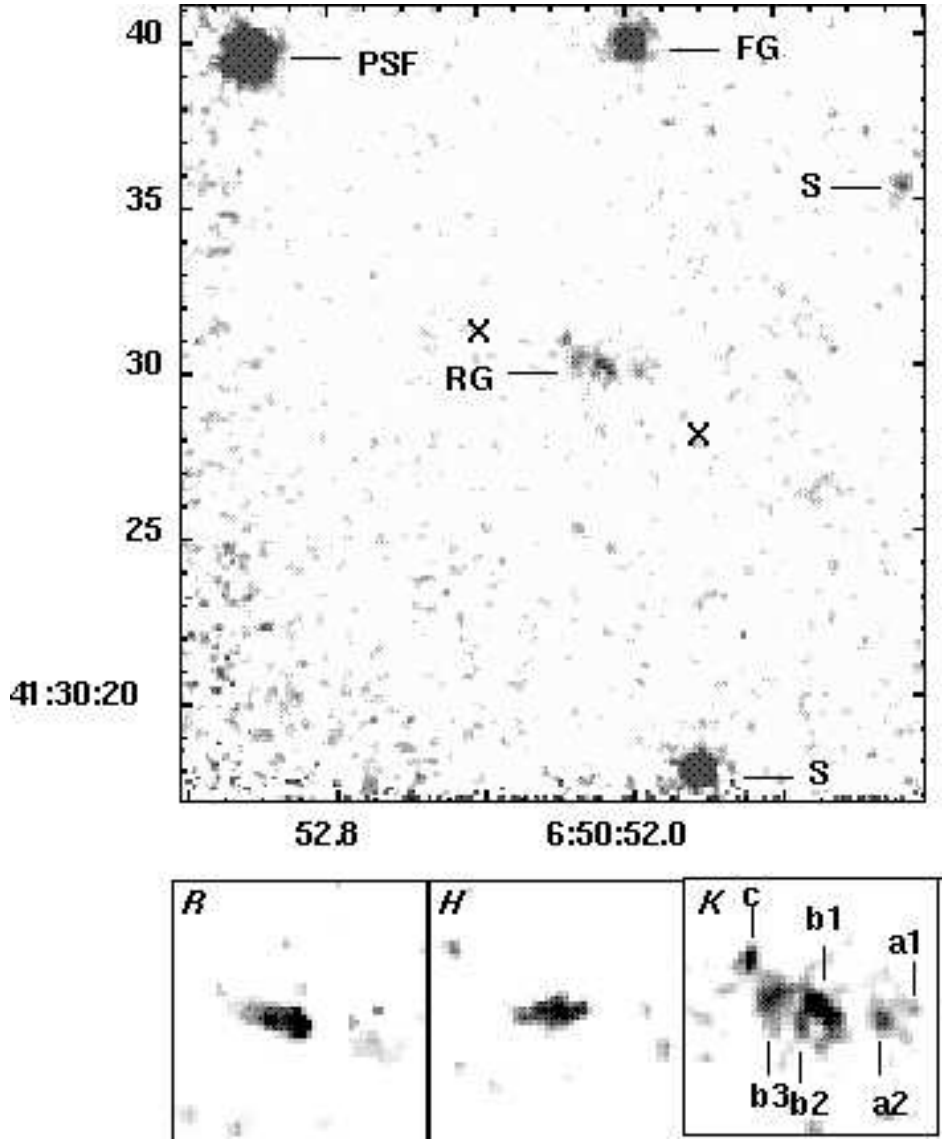}
\caption{Same as Figure~\ref{figure_3c356} for the 4C +41.17 field. In this
case the top panel shows our $K$-band image. See text for details.}
\label{figure_4c+41.17}
\end{figure}

A $K$-band image was obtained by \citet{Chambers1990b} under $\sim2${\arcsec}
seeing but showed only that the RG was elongated.
Deeper NIR images were obtained by \citet{Graham1994} using the Keck Telescope. These latter images achieved a resolution of 0.65{\arcsec} at $K$ and resolved the RG into
two clumps which are aligned along the radio axis.  
Our $K$ image of the RG is very complex (top panel of Figure~\ref{figure_4c+41.17}). 
It is composed of 6 distinct structures which are compact knots. The 
morphology of the RG in $H$ is less complicated. It appears as an elongated
structure running east-west, somewhat misaligned with main axis of the
$K$ band image, which runs from northeast to southwest. A compact unresolved
radio core corresponds to a position 0.7{\arcsec} west of the brightest
emission in $H$ - knot `b1'. \citet{Chambers1996} quote an
internal positional uncertainty of 0.5{\arcsec} so it is possible that the radio
core corresponds to this rest-frame optical knot.  

Pre-refurbishment optical HST images were obtained by \citet{Miley1992} but
the post-refurbishment WFPC2 data discussed here were obtained by \citet{vanBreugel1998}. The close correspondence of the optical and
radio structures leads van Breugel et al. to suggest that star formation 
has been induced by passage of the northeastern radio jet.
The $R$-band image is displayed along with our $H$ and $K$ images in the bottom three panels of Figure~\ref{figure_4c+41.17}. The morphology of the 
RG in $R$ is less complicated than in $K$ and composed of 3 knots which do not exactly register with the knots in the $K$ image. This is not surprising because, in the rest-frame, the 4000{\AA} break separates $R$ and $K$. The
mechanism responsible for the striking morphology in $K$ is one with emission
dominant at rest-frame-visible wavelengths. Since the elongation of the $K$
image is so closely aligned with the radio axis a plausible source is
jet-induced star formation. 

Using deep Keck Telescope spectropolarimetry observations, \citet{Dey1997} find
that the rest-frame UV continuum radiation is unpolarized (a 2-$\sigma$ limit of $P<2.4$\%). This suggests that scattered light from an obscured AGN is not to be
likely the source of this emission. Furthermore, the absorption line UV spectrum
is similar to that of star-forming regions in nearby galaxies. This, along with the low polarization, suggests that the UV emission is due to star formation.

\section{Analysis}\label{analysis}

The goal of comparing our NIR data to the existing WFPC2 images was to
determine if the complicated morphology
exhibited by our sample of RGs in the rest-frame UV persisted at visible wavelengths and if the colors of those sub-components are consistent with mature stellar populations.
A simple analysis by eye of the images in Figures~\ref{figure_3c356}, \ref{figure_3c230}, \ref{figure_3c68.2}, \ref{figure_3c294}, \ref{figure_txs0828+193}, and \ref{figure_4c+41.17} reveals that the RGs are 
complex objects at rest-frame visible wavelengths; moreso for those with redshifts of $z>2$. That is, they
appear to be composed of knots of roughly equal
brightness in the rest-frame visible bandpass. All of the galaxies would be classified as irregulars in both rest-frame UV and visible
light in the standard Hubble sequence.
The absolute brightness of the RGs, however, is roughly constant with redshift and consistent with values for giant elliptical galaxies at low redshift \citep{McCarthy1993}. 

\subsection{Stellar Populations}\label{stellar}

The spectral synthesis models of \citet{Bruzual1993} were used to
estimate the ages of potential stellar populations in the RGs. These yield the Galaxy Isochrone Synthesis Spectral Evolution Library
(GISSEL) isochrones. By experiment with the parameters of their code it was
determined that very simple models of star formation and galaxy evolution were
sufficient to match the observed colors of the objects in the fields within the
errors.

The GISSEL approach is to start with an initial mass function (IMF),
star-formation rate, and a set of stellar evolutionary tracks to predict the
evolution of a population of stars. At each time-step for a given mass of star the position along its
stellar evolutionary track is determined. A library of stellar spectra is
available which represents increments along the stellar evolution tracks. The
composite spectrum of the evolving population of stars is calculated by adding
the spectra of the individual stars from that library and weighting these
according to the IMF. By convolving the resulting spectral energy distribution
(SED) with functions for reddening and observational passbands observed galaxy
colors can be predicted.

\begin{figure}
\plotone{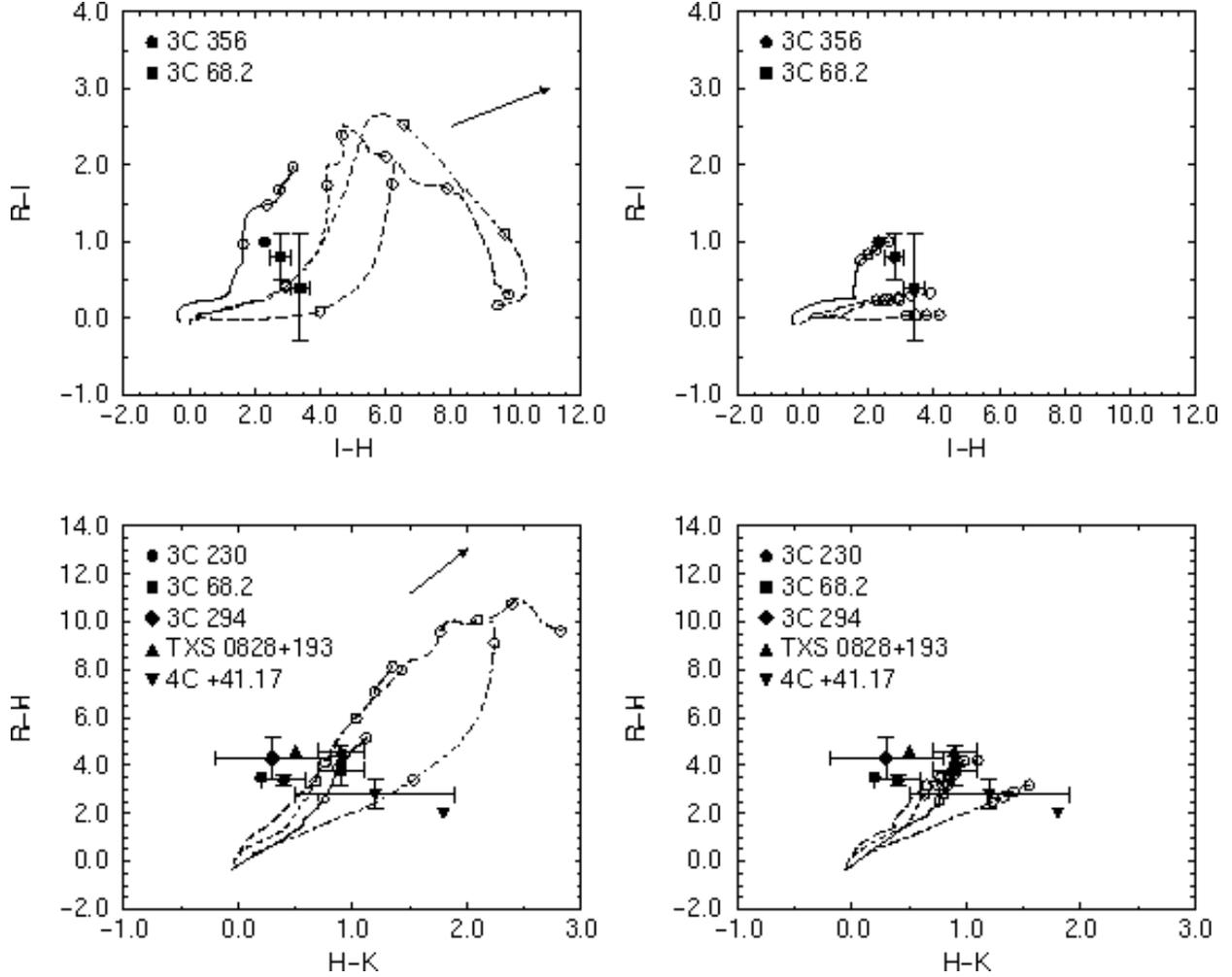}
\caption{The $R-I$ versus $I-H$ (top) and $R-H$ versus $H-K$ (bottom)
color-color plots for all the RG knots. Error bars are given only for the
dominant component for each RG. The overplotted lines are the
galaxy isochrones described in the text. The isochrone for a galaxy at $z=1$ is
indicated by a solid line, $z=2$ by a short-dashed line, $z=3$ by a long-dashed line, and $z=4$ by a dot-dashed line.
The epoch is indicated by open circles
at 1, 2, 5, and 20 Gyr after the onset of star formation. The left-hand panels represent a model with a single burst of star formation of 1 Gyr duration. A vector representing the shift in the models associated with interal reddening of $A_V=2.0$ is given.
A model of continuous star fomation is shown in the right-hand panels.}
\label{figure_colors}
\end{figure}

In Figure~\ref{figure_colors} the
$R-I$ versus $I-H$ and $R-H$ versus $H-K$ colors of the RG components 
are overplotted
with isochrone predictions corresponding to three main models. The first model
is a single burst of star formation lasting 1 Gyr followed by passive stellar
evolution. The second is the same as the first, except that strong reddening,
$A_V=2.0$, is introduced. This is due to the evidence for obscuration in $R-K$
colors, specifically in \objectname{3C 230}, \objectname{3C 294}, and
\objectname{TXS 0828+193}. The third
model is one of continuous star formation at a rate of 
100 $M_{\odot}$ ${\rm yr}^{-1}$. This high rate of star formation was
purposefully chosen to `saturate' the blue colors. That is, a higher rate of
star formation cannot make the measured colors of the galaxy any bluer. These
particular models were chosen because they bracket the extremes of expectations
for evolution for these galaxies. In general the isochrones from the reddened
passive evolution model are too red to match the observed colors. The
predictions from the continuous star formation model are, likewise, too blue.
In each plot a solid line indicates the isochrone for a galaxy at $z=1$. That is, it is a track giving the expected {\it observed} color of a galaxy at $z=1$
given the chosen model evolution. The epoch is indicated by open circles at
1, 2, 5, and 20 Gyr after the onset of star formation. A short-dashed line
indicates the isochrone for the same model galaxy if it were at $z=2$, a long-dashed line for $z=3$, and a dot-dashed line for $z=4$.
The GISSEL models suggest that the colors of all the RGs
in the sample are blue and well fitted by a young or bursting stellar
population. In the case of \objectname{3C 356} some reddening would be necessary to explain
the $R-I$ versus $I-H$ colors of the RG. In all cases the errors in the color
measurements make it difficult to make any strong distinction between ages of
the individual components of the RG. All of the color
information is consistent with stellar populations younger than 5 Gyr for all of
the RGs.

\subsection{Model Galaxy Profiles}\label{profiles}

The previous analysis suggests that the RGs are inconsistent with a population
of relaxed evolved elliptical galaxies at these redshifts. It is interesting to
consider if the removal of the knots from the images might reveal an underlying
galaxy which is consistent with this picture. To investigate this
possibility we developed a method to model the profiles of the galaxies and
subtract the knots.

For each RG a 4{\arcsec} $\times$ 4{\arcsec} image was
extracted for each filter and rotated to a common alignment along the axis of
the greatest elongation. For each image the pixel values were summed along a
direction perpendicular to this axis. In effect, the images were compressed to
slices along the axis of the RG. These profiles were normalized to have peak
fluxes of unity. 

We hypothesize that the galaxies are composed entirely
of bright unresolved regions of young stars. Thus a model of each RG for each
filter was generated as a composition of Gaussian profiles. The Gaussian had a
FWHM given by the image resolution in Table~\ref{table_image_resolution}. The
model galaxy was built up
by starting with a blank image the same size as the RG image and adding a
Gaussian with peak flux and position determined from the photometry. This 
image was then summed in the same manner as for the real data and
normalized to a peak flux of unity.

\begin{figure}
\plotone{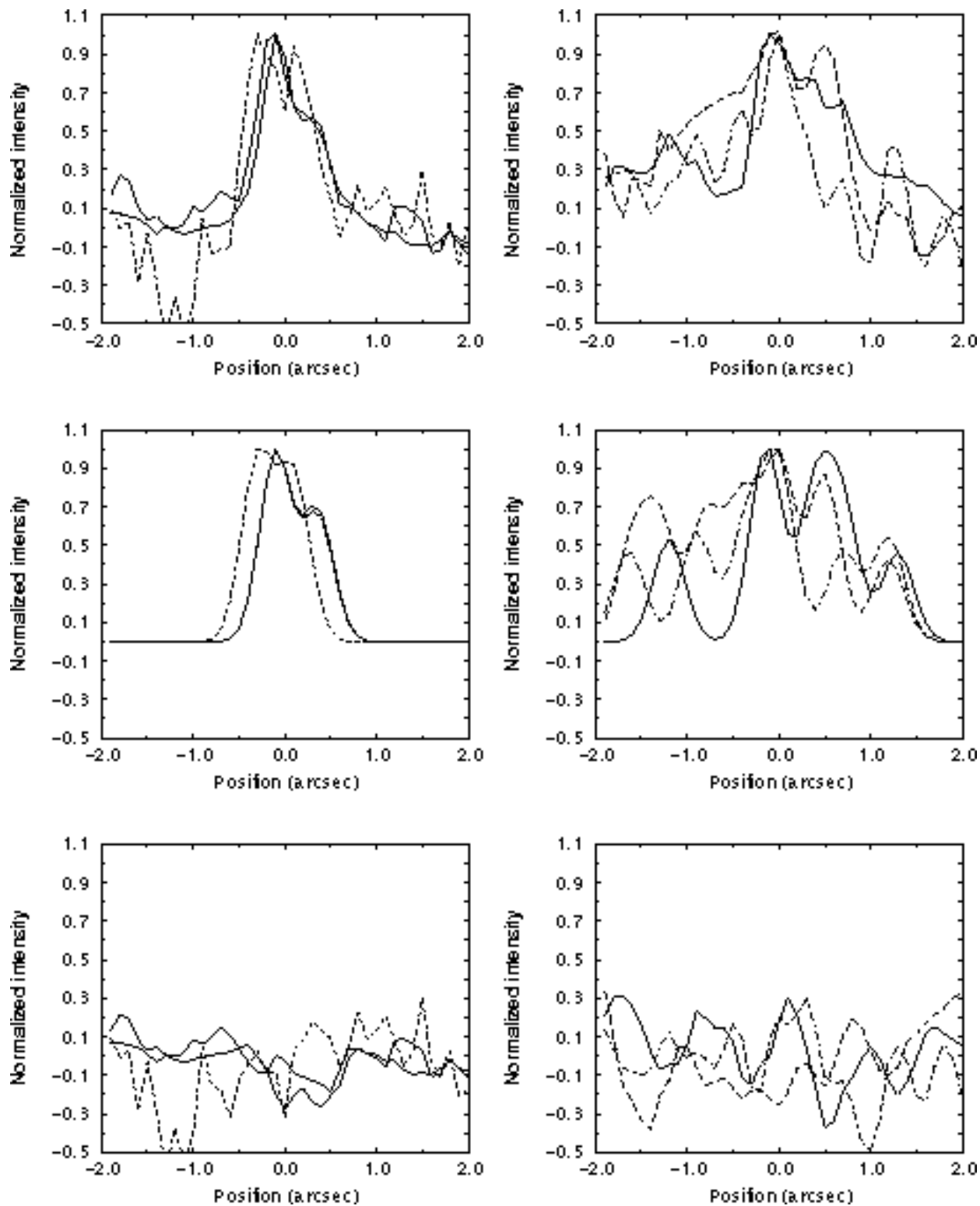}
\caption{The top panels are profiles of the RG for 3C 356 (left) and
4C +41.17 (right). The $R$-band data is represented by a thin solid line, $I$
by thick solid line, $H$ by a long-dashed line, and $K$ by a dot-dashed line.
The models for 3C 356 and 4C +41.17 are shown in the centre panels and the
residuals are shown in the bottom panels.}
\label{figure_slices}
\end{figure}

For each RG and each filter the model galaxy was subtracted from the data. The
results for \objectname{3C 356} and \objectname{4C +41.17} will serve as examples, and they are shown in 
Figure~\ref{figure_slices}. The data are shown in the top panels, the
models in the centre panels, and the residuals in the bottom panels. These
plots show that the model
galaxies account for almost all of the light in the data. In some cases the
residuals show a sharp peak (for example at $+0.2${\arcsec} in $R$ for 
4C +41.17) but this peak is always of the same FWHM as the imaging resolution 
for that bandpass. This can only suggest that an unresolved knot was missed by the method of Section~\ref{photometry}. Accounting for these missed knots leaves residuals in each
band for all the RGs that are random and have $\sigma<\pm0.3$ in normalized flux
to the peak of the original image. This is consistent with the noise in the
images.

The net result, then, is that a model of unresolved knots can account for 
all of the flux in the RG images. We point out that the detection limit for point sources in our AO imaging is approximately only 21 magnitudes in $H$. These limits are even poorer for low-surface-brightness distributions, aggravated by the small angular sizes of the pixels in MONICA and KIR.
So we cannot say that a low-surface-brightness
component does not exist, but to the detection limit of the imaging none of the
galaxies can be fitted with a de Vaucoleurs' $r^{1/4}$ profile. 

\section{Conclusions}\label{conclusions}

None of the RGs look like a typical low redshift RG. Both the
alignment effect and the irregular morphology exhibited at rest-frame
UV wavelengths persist into the visible for this sample. We see no
evidence of underlying relaxed galaxies composed of mature stars, although the
low $S/N$ of these observations does not exclude them. The colors of
the compact knots seem to indicate stellar populations younger than 5 Gyr 
for all of the RGs, yet the overall brightness of the RGs is consistent with
low redshift giant ellipticals. A plausible explanation is that these objects are being assembled by mergers at high redshift.
 
Certainly this study demonstrates the benefit of obtaining AO NIR
imaging of these systems. This work suggests that a least some
HzRGs are not passively evolved ellipticals at $z>2$, but deeper imaging of
a large sample is needed.
The sample provided by natural guide-stars is insufficient and therefore the
path will be provided by the next generation of AO systems, notably for Gemini and Keck, employing laser beacons.



\clearpage

\begin{deluxetable}{lcccrcccc}
\tablecaption{CFHT PUEO Journal of Observations\label{table_cfht_pueo}}
\tablewidth{0pt}
\rotate
\tablehead{&
&\multicolumn{2}{c}{Guide Star} & & &\multicolumn{3}{c}{Exposure time (s)} \\
\cline{3-4} \cline{7-9}
\colhead{Target} &\colhead{$z$}
&\colhead{Mag. ($R$)} &\colhead{Offset ({\arcsec})} &\colhead{Date} &\colhead{Camera} &\colhead{$J$} &\colhead{$H$} &\colhead{$K$}} 
\startdata
\objectname{3C 356} &1.08 &15.0 &16 
&14, 15, 16 Jun 1997 &MONICA &3600 &14400 &\nodata \\
\objectname{3C 230} &1.49 &14.7 &\phn4
&17 Jan 1998 &KIR &\nodata &\phn1200 &\phn3600 \\
\objectname{3C 68.2} &1.58 &14.2 &14
&17, 18 Jan 1998 &KIR &\nodata &\phn2400 &\phn4800 \\
\objectname{3C 294} &1.78 &11.7 &\phn9
&18 Jan 1998 &KIR &\nodata &\phn2400 &\phn4800 \\
\objectname{TXS 0828+193} &2.57 &12.7 &38
&17 Jan 1998 &KIR &3600 &\phn1200 &\phn6000 \\
\objectname{4C +41.17} &3.80 &14.1 &24
&18 Jan 1998 &KIR &1200 &\phn3600 &10800 \\
\enddata
\end{deluxetable}

\begin{deluxetable}{lcccc} 
\tablecaption{HST WFPC2 Archival Data\label{table_hst_wfpc2}}
\tablewidth{0pt}
\tabletypesize{\small}
\tablehead{&\multicolumn{4}{c}{Exposure time (s)}\\
\cline{2-5}
\colhead{Target} &\colhead{F675W} &\colhead{F702W} &\colhead{F785LP} &\colhead{F814W}}
\startdata
\objectname{3C 356} &\nodata &\phn\phn600 &\nodata &1700\\
\objectname{3C 230} &\nodata &\phn\phn560 &\nodata &\nodata \\
\objectname{3C 68.2} &\nodata &\phn\phn600 &3400 &\nodata \\
\objectname{3C 294} &\nodata &\phn1120 &\nodata &\nodata \\
\objectname{TXS 0828+193} &4000 &\nodata &\nodata &\nodata \\
\objectname{4C +41.17} &\nodata &21600 &\nodata &\nodata \\
\enddata
\end{deluxetable}

\begin{deluxetable}{lccccc}
\tablecaption{Image Resolution\label{table_image_resolution}}
\tablewidth{0pt}
\tabletypesize{\small}
\tablehead{&\multicolumn{5}{c}{FWHM ({\arcsec})}\\
\cline{2-6}
\colhead{Target} &\colhead{$R$} &\colhead{$I$} &\colhead{$J$} &\colhead{$H$} &\colhead{$K$}}
\startdata
\objectname{3C 356} &0.21 &0.21 &\nodata &0.21 &\nodata \\
\objectname{3C 230} &0.20 &\nodata &\nodata &0.24 &0.24\\
\objectname{3C 68.2} &0.21 &0.21 &\nodata &0.23 &0.26\\
\objectname{3C 294} &0.21 &\nodata &\nodata &0.21 &0.24\\
\objectname{TXS 0828+193} &0.23 &\nodata &0.45 &0.35 &0.27\\
\objectname{4C +41.17} &0.23 &\nodata &\nodata &0.25 &0.29\\
\enddata
\end{deluxetable}

\begin{deluxetable}{lcccccccccc}
\tablecaption{Photometry of the Radio Galaxy Fields\label{table_photometry}}
\tablewidth{0pt}
\tabletypesize{\small}
\rotate
\tablehead{\colhead{Object} &\colhead{$R$} &\colhead{$I$} &\colhead{$J$} &\colhead{$H$} &\colhead{$K$} 
&\colhead{$R-I$} &\colhead{$R-H$} &\colhead{$I-H$} &\colhead{$H-K$}
&\colhead{$M_H$}}
\startdata
\sidehead{3C 356}
a1 &$21.2\pm0.2$ &$20.4\pm0.1$ &\nodata &$17.6\pm0.2$ &\nodata &$0.8\pm0.3$ &\nodata &$2.8\pm0.3$ &\nodata &$-26.2\pm0.2$\\
a2 &$22.1\pm0.3$ &$21.1\pm0.2$ &\nodata &$18.8\pm0.5$ &\nodata &$1.0\pm0.5$ &\nodata &$2.3\pm0.7$&\nodata  &$-25.0\pm0.5$\\
b &$21.5\pm0.3$ &$21.3\pm0.2$ &\nodata &$17.4\pm0.2$ &\nodata &$0.2\pm0.5$ &\nodata &$3.9\pm0.4$ &\nodata &$-26.4\pm0.2$\\
\sidehead{3C 230}
a &$21.2\pm0.1$ &\nodata &\nodata &$17.8\pm0.1$ &$17.4\pm0.1$ &\nodata &$3.4\pm0.2$ &\nodata &$0.4\pm0.2$ &$-26.6\pm0.1$\\
b &$22.7\pm0.5$ &\nodata &\nodata &$19.2\pm0.3$ &$19.0\pm0.4$ &\nodata &$3.5\pm0.8$ &\nodata &$0.2\pm0.7$ &$-25.2\pm0.3$\\
c &$22.7\pm0.5$ &\nodata &\nodata &$19.2\pm0.3$ &\nodata &\nodata 
&$3.5\pm0.8$ &\nodata &\nodata &$-25.2\pm0.3$\\
1 &$20.3\pm0.1$ &\nodata &\nodata &$17.6\pm0.1$ &$17.0\pm0.1$ &\nodata &$2.7\pm0.2$ &\nodata &$0.6\pm0.2$ &\nodata \\
\sidehead{3C 68.2}
a &$22.2\pm0.5$ &$21.8\pm0.2$ &\nodata &$18.4\pm0.1$ &$17.7\pm0.1$ &$0.4\pm0.7$ &$3.8\pm0.6$ &$3.4\pm0.3$ &$0.9\pm0.2$ &$-26.3\pm0.1$\\
b &\nodata &$22.7\pm0.4$ &\nodata &$20.8\pm0.6$ &$18.9\pm0.5$ &\nodata &\nodata &$1.9\pm1.0$ &$1.9\pm1.1$ &$-23.9\pm0.6$\\
1 &\nodata &$22.2\pm0.2$ &\nodata &$19.4\pm0.2$ &$19.1\pm0.6$ &\nodata &\nodata &$2.8\pm0.4$ &$0.3\pm0.8$ &\nodata \\
2 &$22.0\pm0.5$ &$21.7\pm0.1$ &\nodata &$19.6\pm0.2$ &$19.0\pm0.6$ &$0.3\pm0.6$ &$2.4\pm0.7$ &$2.1\pm0.3$ &$0.6\pm0.8$ &\nodata \\
3 &$21.9\pm0.5$ &$21.1\pm0.1$ &\nodata &$19.5\pm0.2$ &$18.0\pm0.2$ &$0.8\pm0.6$ &$2.4\pm0.7$ &$1.6\pm0.3$ &$1.5\pm0.4$ &\nodata \\
\sidehead{3C 294}
a &$23.4\pm0.8$ &\nodata &\nodata &$19.0\pm0.1$ &$18.7\pm0.4$ &\nodata &$4.4\pm0.9$ &\nodata &$0.3\pm0.5$ &$-25.8\pm0.1$\\
b &\nodata &\nodata &\nodata &$19.5\pm0.2$ &$18.4\pm0.4$ &\nodata &\nodata &\nodata &$1.1\pm0.6$ &$-25.3\pm0.2$\\
c &\nodata &\nodata &\nodata &$20.2\pm0.2$ &$19.3\pm0.8$ &\nodata &\nodata &\nodata &$0.9\pm1.0$ &$-24.6\pm0.2$\\
d &\nodata &\nodata &\nodata &$20.2\pm0.2$ &\nodata &\nodata &\nodata &\nodata &\nodata &$-24.6\pm0.2$\\*
1 &$24.0\pm0.6$ &\nodata &\nodata &$19.1\pm0.1$ &$18.6\pm0.4$ &\nodata &$4.9\pm0.7$ &\nodata &$0.5\pm0.5$ &\nodata \\
2 &$23.9\pm0.6$ &\nodata &\nodata &$18.9\pm0.1$ &$18.3\pm0.3$ &\nodata &$5.0\pm0.7$ &\nodata &$0.6\pm0.4$ &\nodata \\
\sidehead{TXS 0828+193}
a &$23.2\pm0.1$ &\nodata &$20.0\pm0.1$ &$18.6\pm0.1$ &$17.7\pm0.1$ &\nodata &$4.6\pm0.2$ &\nodata &$0.9\pm0.2$ &$-27.2\pm0.1$\\ 
b &$23.5\pm0.1$ &\nodata &$20.4\pm0.1$ &$18.9\pm0.1$ &$18.4\pm0.2$ &\nodata &$4.6\pm0.2$ &\nodata &$0.5\pm0.3$ &$-26.9\pm0.1$\\
c &$24.3\pm0.2$ &\nodata &$21.2\pm0.3$ &$19.4\pm0.3$ &\nodata &\nodata &$4.9\pm0.5$ &\nodata &\nodata &$-26.4\pm0.3$\\
d &$24.5\pm0.3$ &\nodata &\nodata &$20.2\pm0.8$ &$19.3\pm0.4$ &\nodata &$4.3\pm1.1$ &\nodata &$0.9\pm1.2$ &$-25.6\pm0.8$\\
e &\nodata &\nodata &\nodata &$20.2\pm0.8$ &\nodata &\nodata &\nodata &\nodata &\nodata &$-25.6\pm0.8$\\*
1 &$23.7\pm0.1$ &\nodata &$20.4\pm0.1$ &$18.7\pm0.1$ &$18.0\pm0.2$ &\nodata &$5.0\pm0.2$ &\nodata &$0.7\pm0.3$ &\nodata \\
\sidehead{4C +41.17}
a1 &\nodata &\nodata &\nodata &\nodata &$21.0\pm1.2$ &\nodata &\nodata &\nodata &\nodata &\nodata \\
a2 &$23.8\pm0.3$ &\nodata &\nodata &\nodata &$20.0\pm0.5$ &\nodata &\nodata &\nodata &\nodata &\nodata \\
b1 &$23.0\pm0.1$ &\nodata &\nodata &$20.2\pm0.5$ &$19.0\pm0.2$ &\nodata &$2.8\pm0.6$ &\nodata &$1.2\pm0.7$ &$-26.6\pm0.5$\\
b2 &\nodata &\nodata &\nodata &$21.1\pm1.2$ &$20.2\pm0.5$ &\nodata &\nodata &\nodata &$0.9\pm1.7$ &$-25.7\pm1.2$ \\
b3 &$23.0\pm0.1$ &\nodata &\nodata &$21.0\pm1.2$ &$19.2\pm0.2$ &\nodata &$2.0\pm1.3$ &\nodata &$1.8\pm1.4$ &$-25.8\pm1.2$\\
c &$25.3\pm1.2$ &\nodata &\nodata &\nodata &$19.9\pm0.5$ &\nodata &\nodata &\nodata &\nodata &\nodata \\
\enddata
\end{deluxetable}

\begin{deluxetable}{lccccc}
\tablecaption{Emission-Line Contamination\label{table_emission_line}}
\tablewidth{0pt}
\tabletypesize{\small}
\tablehead{&\multicolumn{5}{c}{Contamination (\%)}\\
\cline{2-6}
\colhead{Target} &\colhead{$R$} &\colhead{$I$} &\colhead{$J$} &\colhead{$H$} &\colhead{$K$}}
\startdata
\objectname{3C 356} &\nodata &18\tablenotemark{c} &\nodata &\nodata &\nodata \\
\objectname{3C 230} &\nodata &\nodata &\nodata &13\tablenotemark{d} &\nodata \\
\objectname{3C 68.2} &\nodata &\phn7\tablenotemark{c} &\nodata &14\tablenotemark{d} &\nodata \\
\objectname{3C 294} &\nodata &\nodata &\nodata &\nodata &\nodata \\
\objectname{TXS 0828+193} &\phn5\tablenotemark{a} &\nodata &\nodata &\nodata &\phn5\tablenotemark{e} \\
\objectname{4C +41.17} &15\tablenotemark{b} &\nodata &\nodata &\nodata &15\tablenotemark{f} \\
\enddata
\tablenotetext{a}{\ion{C}{3}]; \citet{Pentericci1999}}
\tablenotetext{b}{\ion{Si}{4}, \ion{C}{4}; \citet{Dey1997}}
\tablenotetext{c}{[\ion{Ne}{5}], [\ion{O}{2}], [\ion{Ne}{3}]; \citet{Best2000}}
\tablenotetext{d}{H$\alpha$; estimated from composite spectrum}
\tablenotetext{e}{H$\alpha$, [\ion{N}{2}]; \citet{Evans1998}}
\tablenotetext{f}{H$\beta$, [\ion{O}{3}]; \citet{Eales1993}}
\end{deluxetable}

\clearpage

\end{document}